\documentclass[conference,10pt]{IEEEtran}
\usepackage{graphicx,cite,amssymb,amsmath,bm}
\usepackage{mathrsfs}
\usepackage[pagewise]{lineno}
\usepackage{color,soul}
\usepackage{tabulary}
\usepackage{multirow,multicol}
\usepackage{cases}
\usepackage{stfloats}
\usepackage{url}
\usepackage{booktabs}
\usepackage{algpseudocode}
\usepackage{algorithm,algpseudocode}
\usepackage{algorithmicx}
\usepackage[normalem]{ulem}
\ifCLASSOPTIONcompsoc
    \usepackage[caption=false, font=normalsize, labelfont=sf, textfont=sf, subrefformat=parens, labelformat=parens]{subfig}
\else
\usepackage[caption=false, font=footnotesize, subrefformat=parens, labelformat=parens]{subfig}
\fi

\usepackage{amsmath,amssymb}
\usepackage{mathtools} 
\usepackage{xspace}

\newcommand{\herm}{^\text{H}}

\newcommand{\trans}{^\text{T}}

\newcommand{\bx}{\mathbf{x}}
\newcommand{\bz}{\mathbf{z}}

\newcommand{\bs}{\mathbf{s}}

\newcommand{\bg}{\mathbf{g}}

\newcommand{\bw}{\mathbf{w}}
\newcommand{\bY}{\mathbf{Y}}
\newcommand{\by}{\mathbf{y}}
\newcommand{\bI}{\mathbf{I}}

\newcommand{\ba}{\mathbf{a}}
\newcommand{\bb}{\mathbf{b}}

\newcommand{\bv}{\mathbf{v}}

\newcommand{\bu}{\mathbf{u}}

\newcommand{\bOmega}{\boldsymbol{\Omega}}

\newcommand{\bvphi}{\boldsymbol{\varphi}}

\newcommand{\bzero}{\boldsymbol{0}}

\newcommand{\CN}{\mathcal{CN}}
\newcommand{\tc}{\tau_{\textsc{c}}}

\newcommand{\boundellipse}[3]% center, x rad, y rad
{(#1) ellipse [x radius=#2,y radius=#3]
}

\DeclareMathOperator{\E}{\mathsf{E}}

\newcommand{\EX}[1]{\mathsf{E}\left\{{#1}\right\}}

\newcommand{\C}{\mathbb{C}}

\newcommand{\Pp}{\rho_{\mathrm{p}}}

\newcommand{\Pd}{\rho_{\mathrm{d}}}

\newcommand{\tauc}{\tau_\mathrm{c}}

\newcommand{\norm}[1]{{ \left\Vert #1 \right\Vert }}

\newcommand{\tp}{\tau_{\mathrm{p}}}

\newcommand{\euler}{\mathrm{e}}

\makeatletter
\def\@setsize#1#2#3#4{
    \@nomath#1
    \let\@currsize#1
    \baselineskip #2
    \baselineskip \baselinestretch\baselineskip
    \parskip \baselinestretch\parskip
    \setbox\strutbox \hbox{
        \vrule height.7\baselineskip
            depth.3\baselineskip
            width\z@}
    \skip\footins \baselinestretch\skip\footins
    \normalbaselineskip\baselineskip#3#4}
\makeatother

\makeatletter
\newcommand{\setstretch}[1]{
    \def\baselinestretch{#1}%
    \@currsize
    }
\makeatother



\def\BibTeX{{\rm B\kern-.05em{\sc i\kern-.025em b}\kern-.08em
    T\kern-.1667em\lower.7ex\hbox{E}\kern-.125emX}}

\newcommand{\comment}[1]{}

\begin{document}
%\vspace{-3 cm}
%\linenumbers

\begin{figure*}[t!]
\normalsize
Paper published in the proceedings of IEEE SPAWC 2021 - 22nd IEEE International Workshop on Signal Processing Advances in Wireless Communications. \\
Added to IEEE Xplore: November 15, 2021. DOI: 10.1109/SPAWC51858.2021.9593193. 

\

\textcopyright~2021 IEEE. Personal use of this material is permitted.  Permission from IEEE must be obtained for all other uses, in any current or future media, including reprinting/republishing this material for advertising or promotional purposes, creating new collective works, for resale or redistribution to servers or lists, or reuse of any copyrighted component of this work in other works.
\vspace{17cm}
\end{figure*}

\title{Conjugate Beamforming with Fractional-Exponent Normalization and Scalable Power Control in Cell-Free Massive MIMO}
\author{
\IEEEauthorblockN{Giovanni Interdonato and Stefano Buzzi}
\IEEEauthorblockA{Dept. of Electrical and Information Engineering, University of Cassino and Southern Latium, Cassino, Italy \\
\{giovanni.interdonato, buzzi\}@unicas.it}
}

\maketitle

\begin{abstract}
This paper considers a cell-free massive MIMO (CF-mMIMO) system using conjugate beamforming (CB) with fractional-exponent normalization. Assuming independent Rayleigh fading channels,  a generalized closed-form expression for the achievable downlink spectral efficiency is derived, which subsumes, as special cases, the spectral efficiency expressions previously reported for plain CB and its variants, i.e. normalized CB and enhanced CB. 
Downlink power control is also tackled, and a reduced-complexity power allocation strategy is proposed, wherein only one coefficient for access point (AP) is optimized based on the long-term fading realizations. 
Numerical results unveil the performance of CF-mMIMO with CB and fractional-exponent normalization, and show that the proposed power optimization rule incurs a moderate performance loss with respect to the traditional max-min power control rule, but with lower complexity and much smaller overall power consumption.
\end{abstract}


%\section{Introduction} \label{sec:intro}
%\IEEEPARstart{C}{ell-free} massive multiple-input multiple-output (MIMO)~\cite{Ngo2017b,Interdonato2019,Zhang2019b} is a practical and scalable embodiment of network MIMO...
%
%\textbf{Contribution:} In this paper, we propose a generalization of... 
%
%The novelty of this study consists of:
%\begin{itemize}
%\item novelty 1
%\item novelty 2
%\item novelty 3
%\end{itemize}

\section{Introduction}
Cell-free massive MIMO (CF-mMIMO) \cite{Ngo2017b}  is a technology based on a network deployment where several low-hardware-complexity access points (APs), connected by fronthaul links to a central processing unit (CPU), serve, in the same frequency band, a much lower number of user equipments (UEs). The time division duplex (TDD) protocol is used to exploit uplink/downlink channel reciprocity, while uplink channel estimates and downlink precoders can be conveniently computed locally at each AP, so as to alleviate the fronthaul burden. Local computation of the beamformers has made very popular the use of conjugate beamforming (CB) in the design and analysis of CF-mMIMO systems, aided by its analytical simplicity. Recently, variants of the conjugate beamformer, i.e. 
normalized CB (NCB) and enhanced CB (ECB) have been proposed and analyzed in    \cite{Polegre2020, Interdonato2021}; in particular,  results  have shown that ECB provides the highest degree of channel hardening, thus maximizing and stabilizing the system performance. 
Another important aspect of CF-mMIMO systems is the choice of the UE-AP association rule. 
In the original CF-mMIMO formulation \cite{Ngo2017b}, all the UEs in the system were served by all the APs; under optimal minimum-mean-square-error (MMSE) processing this represents the best choice, however, under practical precoding schemes, such approach is no longer optimal; additionally, it is clearly not scalable when the system size grows. To circumvent this problem, a user-centric association rule has been proposed \cite{Buzzi2019c,Ngo2018a}, with each UE being served only by a limited number of APs. %The user-centric strategy achieves a better performance than the full CF-mMIMO system in several scenarios and operating conditions; moreover, it permits reducing the complexity of power control allocation rules. 
The user-centric strategy enables a scalable implementation of a CF-mMIMO system, and permits reducing the complexity of power control allocation rules, at the price of an insignificant performance loss with respect to a full CF-mMIMO system.

This paper considers a CF-mMIMO system and provides two main contributions. First of all, a lower bound to the spectral efficiency of a CF-mMIMO system using CB with fractional-exponent normalization is analytically derived under the assumption of Rayleigh-distributed channel fading and of linear MMSE uplink channel estimation. The new spectral efficiency expression holds for any real-valued exponent and subsumes as special cases the expressions previously found for the case of CB, NCB and ECB. Next, in order to limit the system complexity and increase the system scalability, a reduced complexity  
power control formulation is provided, where the number of variables to be optimized is equal to the number of active APs. Numerical results will reveal the performance of CB with fractional-exponent normalization and will show that the newly scalable power control rule incurs a moderate loss with respect to standard rules while achieving a much lower value of transmit power.

\section{System Model} \label{sec:sysmodel}
We consider a CF-mMIMO system operating in TDD mode, and at sub-6 GHz frequency bands. It consists of $M$ APs, connected through a fronthaul network to a CPU, and equipped with $N$ antennas. The APs are geographically distributed in an area of $D \times D$ squared meters, and coherently serve $K$ single-antenna UEs in the same time-frequency resources.      

The conventional block-fading channel model is considered, according to which the time-frequency grid is arranged in coherence blocks wherein the channel is approximately time-invariant and frequency flat. Each coherence block has length $\tauc = T_{\mathrm{c}} B_{\mathrm{c}}$, where $T_{\mathrm{c}}$ and $B_{\mathrm{c}}$ are the shortest user's coherence time and bandwidth, respectively. Moreover, we assume independent Rayleigh fading channels: letting $\bg_{mk} \in \C^{N}$ be the channel response vector between UE $k$ and AP $m$, then $\bg_{mk} \sim \CN(\bzero,\beta_{mk}\bI_N)$, where $\beta_{mk}$ is the large-scale channel gain resulting from the path loss and correlated shadowing.

In the following, we briefly review the uplink training phase and the signal model for the downlink data transmission phase. 

%Each coherence block must accommodate uplink training (generally pilot-based), uplink and downlink data transmission.\footnote{Downlink training is beneficial in cell-free massive MIMO as shown in~\cite{Interdonato2019b}, but comparable performance gain can be obtained by properly normalizing the CB vector as demonstrated in~\cite{Interdonato2021}.} Since we focus on the downlink, the description of the uplink data transmission phase is omitted. The pilot-based uplink training and downlink data transmission phases follow the conventional operation described in the literature and they are here reported for the sake of completeness. While, our novel generalized characterization starts from~\Secref{sec:GCB}.

\subsection{Uplink Training}
During the uplink training all the $K$ UEs synchronously send a pre-determined pilot sequence whose length in samples is denoted by $\tp$. 
Specifically, let $\sqrt{\tp} \bvphi_k \in \C^{\tp}$ be the pilot sent by the $k$-th UE, with $\norm{\bvphi_k}=1$. The pilots are drawn by a set of $\tau_p$ orthogonal vectors; it may happen that the UEs number $K$ is larger than $\tau_p$, in which case the same pilot can be assigned to more than one UE. Accordingly, the inner product $\bvphi_k\trans \bvphi^\ast_j$ will be one, if UEs $k$ and $j$ share the same pilot sequence, or, alternatively, zero. 

The observable at AP $m$ is thus expressed in matrix form as:
\begin{align}
\bY_{\mathrm{p},m} = \sqrt{\tp \Pp}~\sum\nolimits^K_{k=1} \bg_{mk} \bvphi\trans_k + \bOmega_{\mathrm{p},m} \in \C^{N \times \tp},
\end{align}
where $\Pp$ is the normalized signal-to-noise ratio (SNR) of the uplink pilot symbol, and $\bOmega_{\mathrm{p},m}$ is a matrix of additive noise whose elements are i.i.d.  $\CN(0,1)$. 

Based on a prior and local knowledge of the $\{ \beta_{mk} \}$ coefficients, AP $m$ performs linear MMSE estimation of the $k$-th UE channel 
$\bg_{mk}$ through the processing
\begin{align}
\hat{\bg}_{mk} &=  \underbracket[.5pt]{\frac{\sqrt{\tp\Pp} \beta_{mk}}{\tp\Pp \sum\nolimits^K_{j =1} \beta_{mj} |\bvphi_k\herm \bvphi_j|^2 + 1}}_{\textstyle c_{mk}} \underbracket[.5pt]{\vphantom{\frac{\sqrt{\tp\Pp} \beta_{mk}}{\tp\Pp \sum\nolimits^K_{j =1} \beta_{mj} |\bvphi_k\herm \bvphi_j|^2 + 1}}\bY_{\mathrm{p},m}\bvphi^\ast_k}_{\textstyle \by_{\mathrm{p},mk}}.
\end{align}
Note that $\hat{\bg}_{mk}\!\sim\!\CN(\bzero,\gamma_{mk}\bI_N)$, with $\gamma_{mk}\!=\!\sqrt{\tp\Pp} c_{mk} \beta_{mk}$.
%\begin{equation}\label{eq:gamma}
%\gamma_{mk} = \frac{{\tp\Pp} \beta^2_{mk}}{\tp\Pp \sum\nolimits^K_{j =1} \beta_{mj} |\bvphi_k\herm \bvphi_j|^2 + 1}.
%\end{equation}
%If UE $k$ and UE $j$ share the same pilot, 
For any pair of UEs $k$ and $j$ sharing the same pilot, then $\bvphi_k\herm \bvphi_j = 1$ and $\by_{\mathrm{p},mk} = \by_{\mathrm{p},mj}$. In this case, it holds that
\begin{align} 
\hat{\bg}_{mk} = \dfrac{\beta_{mk}}{\beta_{mj}} \hat{\bg}_{mj},  \qquad
\gamma_{mk} = \dfrac{\beta^2_{mk}}{\beta^2_{mj}} \gamma_{mj} \label{eq:correlated-estimates}.
\end{align}
The channel estimation error is $\tilde{\bg}_{mk} = \bg_{mk} - \hat{\bg}_{mk},$ distributed as $\tilde{\bg}_{mk} \sim \CN(\bzero,(\beta_{mk}-\gamma_{mk})\bI_N),$ and independent of $\hat{\bg}_{mk}$. 

\subsection{User-Centric Downlink Data Transmission}
In order to make the system scalable, each UE is served by a limited number of APs and not by all the APs in the system, i.e. a user-centric AP-UE association is performed \cite{Buzzi2019c}. 
User-centric clusters can be formed in several ways, and we adopt a simple criterion 
according to which the generic $k$-th UE is served by a pre-fixed number of APs, 
i.e. the ones with the largest large-scale channel gain towards UE $k$.
We denote by  $\mathcal{M}_k$ the set of APs serving user $k$. Similarly,  $\mathcal{K}_m$ is used to denote the set of UEs served by AP $m$.

Let $\bw_{mk} \in \C^N$ be the precoding vector used by AP $m$ in the service of UE $k$, then the precoded data symbol sent to the desired UEs is given by 
\begin{equation} \label{eq:data-transmission}
\bx_m = \sqrt{\Pd} \sum\limits_{k \in \mathcal{K}_m} \sqrt{\eta_{mk}} \bw_{mk} q_k,
\end{equation}
where $q_k$ is the uncorrelated data symbol intended for UE $k$, with $\EX{|q_k|^2} = 1$. 
The maximum downlink normalized SNR is denoted by $\Pd$, and $\{\eta_{mk}\}$ are power control coefficients satisfying the per-AP power constraint
\begin{equation}\label{eq:power-constraint}
\EX{\norm{\bx_m}^2} \leq \Pd, \; m=1,\ldots,M.
\end{equation}
The received signal at UE $k$ resulting from the joint coherent transmission of the $M$ APs is thus given by
\begin{align} \label{eq:data-symbol}
r_k % &= \sum\limits^M_{m =1} \bg\trans_{mk} \bx_m + \omega_k \nonumber \\
&= \sqrt{\Pd} \sum^K\limits_{j = 1} \sum\limits_{m \in \mathcal{M}_j}  \sqrt{\eta_{mj}} \bg\trans_{mk} \bw_{mj} q_j + \omega_k,  
\end{align}
where $\omega_k\sim\CN(0,1)$ is additive noise. 

%\begin{figure*}[!b]
%\normalsize
%\setcounter{eqcnt1}{\value{equation}}
%\setcounter{equation}{17}
%\hrulefill
%\begin{equation}
%\label{eq:SINR:CB}
%\mathsf{SINR}\CB_k = \frac{\Pd N^2 \left( \sum\limits_{m=1}^M \sqrt{\eta_{mk}} \gamma_{mk} \right)^2}{\Pd N \sum\limits_{j = 1}^K \varsigma_{kj} + \Pd N^2 \sum\limits_{j \neq k}^K \left(\sum\limits_{m=1}^M \sqrt{\eta_{mj}} \gamma_{mj} \dfrac{\beta_{mk}}{\beta_{mj}} \right)^2\!\left|\bvphi_k\herm\bvphi_j\right|^2 + 1},
%\end{equation}
%\hrule
%\vspace*{4pt}
%\setcounter{equation}{20}
%\begin{equation} \label{eq:SINR:NCB}
%\mathsf{SINR}\NCB_k = \frac{\Pd \alpha^2 \left(\sum\limits_{m=1}^M \sqrt{\eta_{mk} \gamma_{mk}}\right)^2}{\Pd (N-1-\alpha^2) \sum\limits_{m=1}^M \eta_{mk} \gamma_{mk}+\Pd \sum\limits_{j=1}^K \sum\limits_{m=1}^M \eta_{mj} \beta_{mk}+\Pd \sum\limits_{j \neq k}^K \Upsilon_{kj}~|\bvphi\herm_k \bvphi_j|^2+1},
%\end{equation}
%where
%\begin{align}
%\Upsilon_{kj} &\triangleq (N-1) \sum\limits_{m=1}^M \eta_{mj} \gamma_{mj} \frac{\beta^2_{mk}}{\beta^2_{mj}} + \alpha^2 \sum\limits_{m=1}^M \sum\limits_{n \neq m}^M \sqrt{\eta_{mj}\eta_{nj}\gamma_{mj}\gamma_{nj}}\frac{\beta_{mk}\beta_{nk}}{\beta_{mj}\beta_{nj}}, \label{eq:NCB:upsilon}\\
%\alpha &\triangleq \frac{\Gamma{(N+1/2)}}{\Gamma{(N)}}. \label{eq:alpha}
%\end{align}
%\setcounter{equation}{\value{eqcnt1}}
%\end{figure*}

\section{Conjugate Beamforming with Fractional-Exponent Normalization} \label{sec:GCB}

The need to confine to the AP the beamformer computation, in order to alleviate the fronthaul burden, as well as their convenient mathematical tractability, has made CB schemes extremely popular for CF-mMIMO. Classical CB, studied in  \cite{Ngo2018a}, is obtained by setting $\bw_{mk} = \hat{\bg}^\ast_{mk}$. CB poorly contributes to make the effective downlink channel gain nearly deterministic as demonstrated in~\cite{ZChen2018}; to circumvent this problem, normalized CB (NCB), consisting in setting $\bw_{mk} = {\hat{\bg}^\ast_{mk}}/{\norm{\hat{\bg}_{mk}}}$, and enhanced normalized CB (ECB), which corresponds to $\bw_{mk} = {\hat{\bg}^\ast_{mk}}/{\norm{\hat{\bg}_{mk}}}^2$ were 
proposed and analyzed in~\cite{Polegre2020} and in~\cite{Interdonato2021}, respectively. {\footnote{The modified CB in~\cite{Attarifar2019} requires instead CSI exchange among the APs.} In this paper, we consider a general CB with fractional-exponent normalization, i.e. we let \begin{equation} \label{eq:GCB:precoding-vector}
\bw_{mk} = \frac{\hat{\bg}^\ast_{mk}}{\norm{\hat{\bg}_{mk}}^{\alpha+1}},
\end{equation}
with $\alpha$, the \textit{channel inversion rate}, an arbitrary real-valued parameter.
Clearly, the conventional CB, NCB and ECB are special cases of the proposed CB scheme, obtained for  $\alpha = -1, \, 0, \, 1$,  respectively.

Let us now concentrate on the analysis of the proposed CB scheme with fractional-exponent normalization.  
By inserting~\eqref{eq:GCB:precoding-vector} into~\eqref{eq:data-transmission} and working out the expectation in~\eqref{eq:power-constraint}, it results that the power control coefficients $\{\eta_{mk}\}$ must satisfy the per-AP power constraint
\begin{equation} \label{eq:GCB:power-constraint}
\frac{\Gamma(N-\alpha)}{\Gamma(N)} \sum\limits_{k \in \mathcal{K}_m} \frac{\eta_{mk}}{\gamma^{\alpha}_{mk}}  \leq 1, \; m = 1, \ldots, M,
\end{equation}
where $\Gamma(\cdot)$ is the \textit{Gamma} function. (See proof in Appendix~\textit{A}.)

Under the assumption of Rayleigh fading, a lower bound to the achievable SE can be found by
using the popular  \textit{hardening} lower bound~\cite{massivemimobook},  by treating all the interference and noise contributions in~\eqref{eq:data-symbol} as uncorrelated effective noise: 
\begin{equation} \label{eq:SE}
\mathsf{SE}_k \!=\! \bar{\xi} \log_2 \!\left(\!1\!+\!\frac{|\mathsf{DS}_k|^2}{\EX{|\mathsf{BU}_k|^2}\!+\!\sum\nolimits^K_{j \neq k} \EX{|\mathsf{UI}_{kj}|^2}\!+\!1}\!\right)\!,
\end{equation}
%\left(1 - \frac{\tp}{\tauc} \right)
%\begin{align} \label{eq:data-symbol-capacity}
%r_k = \mathsf{DS}_k q_k + \mathsf{BU}_k q_k + \sum\limits^K_{j \neq k} \mathsf{UI}_{kj} q_j + \omega_k,
%\end{align}
where the pre-log factor $\bar{\xi} = \xi \left(1 - {\tp}/{\tauc} \right)$, with $0<\xi<1$, accounts for the share of the coherence block reserved to the downlink and for the pilot overhead. Moreover, we have
%\begin{equation} \label{eq:SINR}
%\mathsf{SINR}_k = \frac{|\mathsf{DS}_k|^2}{\EX{|\mathsf{BU}_k|^2} + \sum\limits^K_{j \neq k} \EX{|\mathsf{UI}_{kj}|^2}  + 1}.
%\end{equation}
\begin{align}
\mathsf{DS}_k \! &= \! \sum\limits_{m \in \mathcal{M}_k}\!\sqrt{\Pd\eta_{mk}}~\EX{\bg\trans_{mk} \bw_{mk}}, \label{eq:DS} \\
\mathsf{BU}_k \! &= \! \sum\limits_{m \in \mathcal{M}_k}\!\sqrt{\Pd\eta_{mk}} \left(\bg\trans_{mk} \bw_{mk}\!-\!\EX{\bg\trans_{mk} \bw_{mk}} \right), \label{eq:BU} \\
\mathsf{UI}_{kj} \! &= \! \sum\limits_{m \in \mathcal{M}_j}\!\sqrt{\Pd\eta_{mj}}~\bg\trans_{mk} \bw_{mj}, \label{eq:UI} 
\end{align}
where, $\mathsf{DS}_k$ is the signal desired by UE $k$, $\mathsf{BU}_k$ (beamforming gain uncertainty) is a self-interference contribution due to the UE's lack of CSI, and $\mathsf{UI}_{kj}$ represents the inter-user interference. 

Computing all the expectations in~\eqref{eq:SE}, an achievable downlink SE in closed form is given by $\mathsf{SE}_k = \bar{\xi} \log_2(1+ \mathsf{SINR}_k)$, where the effective signal-to-interference-plus-noise ratio (SINR) is equal to 
\begin{equation}
\label{eq:SINR:parametrized}
\mathsf{SINR}_k \!=\! \frac{\left(  \sum\limits_{m \in \mathcal{M}_k} \sqrt{\rho_{mk}} \mathrm{a}_{mkk}\right)^2}{\sum\limits^K_{j =1} \sum\limits_{m \in \mathcal{M}_j}\!\rho_{mj} \mathrm{b}_{mkj}\! + \!\sum\limits^K_{j \neq k}\!\left(\!\sum\limits_{m \in \mathcal{M}_j}\! \sqrt{\rho_{mj}}\mathrm{a}_{mkj} \!\right)^2 \!\!\!+ \!1},
\end{equation}
with $\rho_{mk} = \Pd \eta_{mk}$, and
\begin{align*}
\begin{cases}
\mathrm{a}_{mkj} 
%&= \frac{\Gamma\left(N-\frac{\alpha-1}{2}\right)}{\Gamma(N)} \frac{\beta_{mk}^\alpha}{\beta_{mj}^\alpha} \gamma_{mk}^{\mbox{$\frac{1-\alpha}{2}$}} |\bvphi\herm_k \bvphi_j|, \\
\!\!\!\!&= \dfrac{\Gamma\left(N-\frac{\alpha-1}{2}\right)}{\Gamma(N)} \dfrac{\gamma^{1/2}_{mk}}{\gamma_{mj}^{\alpha/2}} |\bvphi\herm_k \bvphi_j|, \\
\mathrm{b}_{mkj}
%&= \frac{\Gamma(N-\alpha)}{\Gamma(N)} (N\!-\!\alpha\!-\!1) \frac{\beta_{mk}^{2\alpha}}{\beta_{mj}^{2\alpha}}\gamma_{mk}^{1-\alpha}|\bvphi\herm_k \bvphi_j|^2 - \mathrm{a}_{mkj}^2 \\
%&\quad + \frac{\Gamma(N-\alpha)}{\Gamma(N)}\frac{\beta_{mk}}{\gamma^{\alpha}_{mj}} \nonumber \\
\!\!\!\!&= \dfrac{\Gamma(N-\alpha)}{\Gamma(N)} (N\!-\!\alpha\!-\!1) \dfrac{\gamma_{mk}}{\gamma_{mj}^{\alpha}}|\bvphi\herm_k \bvphi_j|^2 - \mathrm{a}_{mkj}^2 \\
&\quad + \dfrac{\Gamma(N-\alpha)}{\Gamma(N)}\dfrac{\beta_{mk}}{\gamma^{\alpha}_{mj}}.
\end{cases}
\end{align*}
\begin{IEEEproof}
See Appendix \textit{B}.
\end{IEEEproof}
%By setting $\alpha = -1$, $\alpha = 0$ and $\alpha = 1$ in~\eqref{eq:SINR:parametrized}, we obtain the closed-form SINR expressions for CB, NCB and ECB given in~\cite{Interdonato2021}, respectively.   

\section{Scalable Power Control} \label{sec:power-control}

As it can be seen by inspecting~\eqref{eq:data-transmission}, power control may be carried out on two different time scales. A fast power control is performed at the small-scale fading time scale by fine-tuning the fractional-exponent $\alpha$ in the precoding vectors. This fast power control is fully distributed and solely function of local CSI.\footnote{The power control in~\cite{Nikbakht2020} occurs on the slow fading time scale, requires CSI exchange among the APs and the fine-tuning of two fractional exponents.}
A slow power control strategy is then performed at the large-scale fading time scale; such strategy has been usually defined by the power control coefficients $\{\eta_{mk}\}$, which are optimized based on the SINR expression~\eqref{eq:SINR:parametrized}, which does not depend on small-scale fast fading coefficients. Clearly, the optimization with respect to the coefficients $\{\eta_{mk}\}$ requires solving an optimization problem with a number of parameters equal to $MK$, in the full cell-free case where all the APs serve all the UEs \cite{Ngo2017b}, or equal to $\sum_{m=1}^M |\mathcal{K}_m| = \sum_{k=1}^K |\mathcal{M}_k| \ll MK$, in the user-centric approach. 

In this paper, we also consider a simplified power optimization rule wherein we let $\eta_{mk}=\eta_m$, $\forall k \in {\cal K}_m$, i.e. we optimize only one parameter per active AP. Clearly, we are making here a suboptimal choice in order to obtain a simpler and more scalable system.

\subsection{Max-Min Fairness Power Control} \label{sec:MMF}
Max-min fairness (MMF) power control maximizes the minimum SE across active UEs. Formally, the problem is formulated as 
%\begin{subequations} 
\begin{align}	\label{Problem:Max-Min:general}
  \mathop {\max_{\{\eta_{mk} \geq 0 \}}} & \quad \min_{k} \mathsf{SINR}_k\, , &% \\
  \text{s.t.} &\;\; \EX{\norm{\bx_m}^2} \leq \Pd, ~\forall m,
\end{align}
%\end{subequations}
which can be rewritten in \textit{epigraph form} as
\begin{subequations} \label{Problem:Max-Min}
\begin{align}	
  \mathop {\text{maximize}}\limits_{\{\eta_{mk}\geq 0 \},~\nu} & \quad \nu  \\
  \text{s.t.} &\quad \norm{\bs_k} \leq \boldsymbol{\ba}_{kk}\trans \bu_k,~\forall k, \label{constraint:SINR-target} \\
  			  %&\quad\eta_{m} \leq \left(\frac{\Gamma(N-\alpha)}{\Gamma(N)}\sum\limits_{k \in \mathcal{K}_m} \frac{1}{\gamma^{\alpha}_{mk}}~\right)^{-1}\!\!\!,~\forall m, \\
  			  &\quad \norm{\hat{\boldsymbol{\gamma}}_m \circ \bar{\bu}_m} \leq \sqrt{\Pd\frac{\Gamma(N)}{\Gamma(N-\alpha)}},~\forall m,
  			  \label{constraint:power}  			  
\end{align}
\end{subequations}
where $\nu$ is a new variable representing the minimum target SINR, the operator $\circ$ denotes the Hadamard product, and:
\begin{itemize}
\item $\ba_{kj} = \left[\mathrm{a}_{{\mathcal{M}_j(1)}kj}, \ldots, \mathrm{a}_{\mathcal{M}_j(|\mathcal{M}_j|)kj}  \right]\trans$;
\item $\bu_k\!=\!\sqrt{\Pd}\left[\sqrt{\eta_{\mathcal{M}_k(1)k}}, \ldots, \sqrt{\eta_{\mathcal{M}_k(|\mathcal{M}_k|)k}}  \right]\trans$;
\item $\bar{\bu}_m\!=\!\sqrt{\Pd}\left[\sqrt{\eta_{m\mathcal{K}_m(1)}}, \ldots, \sqrt{\eta_{m\mathcal{K}_m(|\mathcal{K}_m|)}}  \right]$;
\item $\hat{\boldsymbol{\gamma}}_m = \left[\gamma_{m\mathcal{K}_m(1)}^{-\alpha/2}, \ldots, \gamma_{m\mathcal{K}_m(|\mathcal{K}_m|)}^{-\alpha/2}  \right]$;
\item $\bs_k = \sqrt{\nu}\cdot\left[\bv_{k}\trans \bI_{-k}, \norm{\bb_{k1} \circ \bu_1}, \ldots, \norm{\bb_{kK} \circ \bu_K}, 1  \right]\trans$;
\item $\bv_{k} \triangleq \left[ \ba\trans_{k1} \bu_1, \ldots, \ba\trans_{kK} \bu_K \right]\trans$;
\item $\bI_{-k}$ is a $K\times(K-1)$ matrix obtained from $\bI_k$ with the $k$th column removed;
\item $\bb_{kj} = \left[\sqrt{\mathrm{b}_{\mathcal{M}_j(1)kj}}, \ldots, \sqrt{\mathrm{b}_{\mathcal{M}_j(|\mathcal{M}_j|)kj}}  \right]\trans$. Note that the elements of this vector are strictly positive if $N > \alpha$.
\end{itemize}
Importantly, this problem formulation keeps the same structure of the formulations described in~\cite{Ngo2017b}, and~\cite{Interdonato2021}, which are however special cases obtained for certain values of $\alpha$. Problem~\eqref{Problem:Max-Min} admits global optimal solutions that can be computed by solving a sequence of second-order cone programs \cite{Ngo2017b},~\cite{Interdonato2021}.

The case in which only one power coefficient per active AP is to be optimized, which we denote by MMF-U (uniform), can be simply obtained by specializing \eqref{Problem:Max-Min} with  $\eta_{mk}=\eta_m$, $\forall k \in {\cal K}_m$. In this case, however, note that 
the per-AP power constraint in~\eqref{constraint:power} is replaced by
\begin{equation}
\label{eq:unif-coeff}
\eta_{m} \leq \frac{\Gamma(N)}{\Gamma(N-\alpha)} \frac{1}{\sum\limits_{j \in \mathcal{K}_m} \gamma^{-\alpha}_{mj}}, \forall m,
\end{equation}
which is intended for all the UEs $k$ such that $m \in \mathcal{M}_k$.

\subsection{Maximal-Ratio Power Control} \label{sec:Maximal-ratio}
As an alternative to the centralized MMF power control, we may consider the \textit{maximal-ratio} (MR) power control~\cite{Ngo2017b}, which consists in setting the power control coefficients as
\begin{align}
\eta_{mk} = \frac{\Gamma(N)}{\Gamma(N-\alpha)} \frac{\gamma_{mk}^{\alpha+1}}{\sum\limits_{j \in \mathcal{K}_m} \gamma_{mj}}.
\end{align}
This heuristic power control strategy has been proved to be effective in cell-free massive MIMO, other than being fully scalable and computationally simple. 
%Moreover, maximal-ratio power control enables full-power transmission by all the APs, thereby ignoring the effects of the interference on the performance of the users with poor channel conditions. 
The variant of this power control rule with one coefficient per AP, i.e. MR-U, is attained by imposing~\eqref{eq:unif-coeff} with equality.

\section{Simulation Results} \label{sec:results}
We provide here some results obtained by averaging several snapshots of the network, i.e., many random realizations of AP and UE locations. Each snapshot determines a set of large-scale fading coefficients $\{\beta_{mk}\}$ capturing pathloss and shadow fading defined as
$\beta_{mk} = \mathsf{PL}_{mk} \cdot 10^{~\sigma_\text{sh}q_{mk} / 10},
$
where $\mathsf{PL}_{mk}$ follows the 3GPP Urban Microcell pathloss model 
%defined in \cite[Table B.1.2.1-1]{LTE2017}, 
\cite{Interdonato2021}, and $\sigma_\text{sh}$ is the standard deviation of the correlated log-normal shadow fading, $q_{mk} \sim \mathcal{N}(0,1)$. Unless otherwise stated, we use the same simulation settings as in~\cite{Interdonato2021}, and: $D = 500$ m, $\sigma_\text{sh}=4$~dB, $\xi=0.5$, $|\mathcal{M}_k| = 5~\forall k$, $\tp = K/2$, and $\tc = 200$ samples resulting from a coherence bandwidth of 200 kHz and a coherence time of 1 ms.
The maximum transmit power per AP and per UE is $200$ mW and $100$~mW, respectively, while the noise power is $-92$ dBm. Hence, $\Pd = 115$ dB and $\Pp = 112$ dB.

First of all, we compare in Fig.  1 the performance of MR and MR-U power control rules, versus the channel inversion rate $\alpha$, focusing on the 95\%-likely, 50\%-likely, and 10\%-likely SE values. Results show that for the worst UEs the optimal value of $\alpha$ is slightly smaller than 0, whereas for the best UEs it is better to have $\alpha=1$, i.e. the ECB.
\begin{figure}[!t] \label{fig:fig1}
\centering
\includegraphics[width=\linewidth]{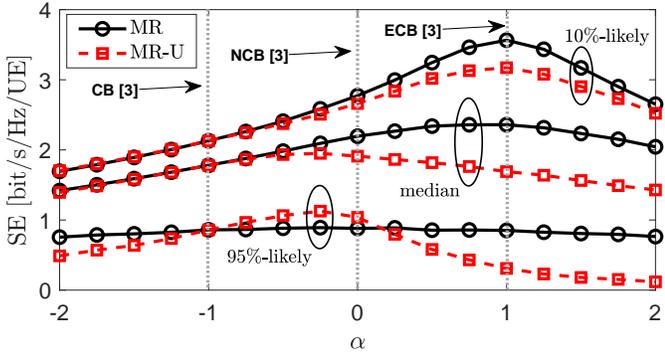} \vspace*{-7mm}
\caption{Achievable SE per UE, for different percentiles, as the channel inversion rate $\alpha$ varies. Here, $M = 200$, $N = 8$, $K = 40$ and maximal-ratio (MR) power control is applied. MR and MR-U coincide when $\alpha = -1$.  }
\end{figure} 
Next, Figs. 2 and 3 are devoted to the comparison of the MR, MMF and MMF-U rules, for several values of $\alpha$, and in terms of average minimum SE and average transmit powers, respectively. Inspecting the figures, it can be seen that the scalable power control rule MMF-U attains a performance in between that of MR and of MMF, while achieving significant savings in terms of transmitted power per active AP.

\section{Conclusion}
With regard to a CF-mMIMO system, this paper has provided two main contributions. First of all, a closed form expression of the SE for CB with fractional-exponent normalization has been proposed, subsuming previously derived SE expressions for other types of conjugate beamformers, and unveiling the system performance for arbitrary values of the normalization exponent. Next, a simplified power control rule has been proposed,  that has been shown to be a good compromise, in terms of scalability, complexity and performance, between local rules and centralized optimized rules, while achieving significant savings in terms of transmit power.

\section*{Acknowledgement}
This paper has been supported by the MIUR-funded PRIN 2017 Project LiquidEDGE.

\begin{figure}[!t]
\centering
\includegraphics[width=.9\linewidth]{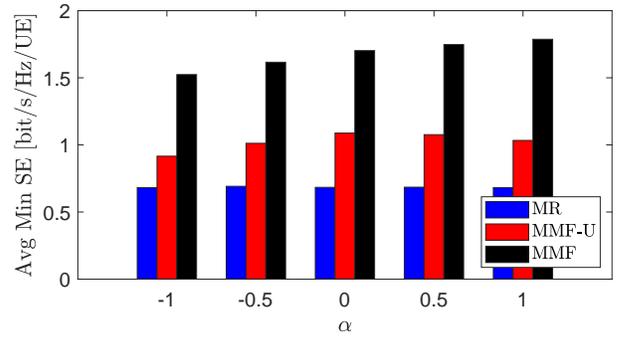} \vspace*{-4mm}
\caption{Average minimum SE per UE for different values of $\alpha$. Here, $M = 100$, $N = 8$, $K = 20$.}
\label{fig:fig2}
\end{figure} 
\begin{figure}[!t]
\centering
\includegraphics[width=.9\linewidth]{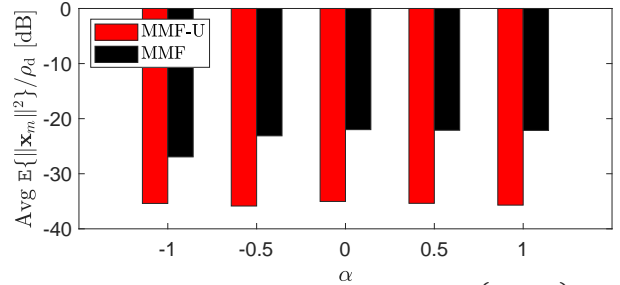} \vspace*{-4mm}
\caption{Average normalized transmit power per AP, $\EX{\norm{\bx_m}^2}/ \Pd$,  for different values of $\alpha$. Here, $M = 100$, $N = 8$, $K = 20$.}
\label{fig:fig3}
\end{figure}

\appendix
\subsection{Proof of~\eqref{eq:GCB:power-constraint}}
The transmit power of the downlink data signal is given by
\begin{equation} \label{eq:transmit-power}
\EX{\norm{\bx_m}^2} = \Pd \sum\limits^K_{k=1} \eta_{mk} \EX{\frac{1}{\norm{\hat{\bg}_{mk}}^{2\alpha}}},~m=1,\ldots,M.
\end{equation}
In order to obtain~\eqref{eq:GCB:power-constraint}, we first need to calculate the expectation in~\eqref{eq:transmit-power} in closed form.
Since $\hat{\bg}_{mk}\!\sim\!\CN(\bzero,\gamma_{mk}\bI_N)$, then $\norm{\hat{\bg}_{mk}}^2 = \dfrac{\gamma_{mk}}{2} \norm{{\bz}}^2$ where ${\bz}\!\sim\!\mathcal{N}(\bzero,\bI_{2N})$, and $ \norm{{\bz}}^2 \sim \chi^2_{2N} $, i.e., follows a central \textit{chi}-square distribution with $2N$ degrees of freedom, and probability density function (pdf)
\begin{align}
f_{\norm{{\bz}}^2}(x) = \frac{x^{N-1} \euler^{-x/2}}{2^N \Gamma(N)}u(x),
\end{align}
where $u(x)$ is the \textit{unit-step} function, and $\Gamma(\cdot)$ is the \textit{Gamma} function. 
%defined as
%\begin{align} \label{eq:gamma-function}
%\Gamma(x) \triangleq \int\nolimits^{\infty}_0 t^{x-1} \euler^{-t} dt.
%\end{align}
Since it holds that $ f_{cX}(x) = \dfrac{1}{c} f_X \left(\dfrac{x}{c}\right),$ we have
\begin{align}
f_{\norm{\hat{\bg}_{mk}}^2}(x) = \frac{x^{N-1} \euler^{-x/\gamma_{mk}}}{\gamma^N_{mk} \Gamma(N)}u(x).
\end{align}
By using the \textit{Law of the unconscious statistician} then
\begin{align} \label{eq:expvalue-alpha}
&\EX{\frac{1}{\left(\norm{\hat{\bg}_{mk}}^2\right)^{\alpha}}} \!= \!\!\int^{\infty}_0 \frac{1}{x^{\alpha}} \frac{x^{N-1} \euler^{-x/\gamma_{mk}}}{\gamma^N_{mk} \Gamma(N)} dx\! =\!\!\!\! %\nonumber \\
%&\quad=\frac{1}{\gamma^N_{mk} \Gamma(N)} \int^{\infty}_0 x^{N-1-\alpha} \euler^{-x/\gamma_{mk}} dx \nonumber \\
&%\quad\stackrel{(a)}{=}
%=\frac{1}{\gamma^{\alpha}_{mk} \Gamma(N)} \! \int^{\infty}_0 t^{N-1-\alpha} \euler^{-t} dt 
\frac{\Gamma(N\!-\!\alpha)}{\gamma^{\alpha}_{mk}\Gamma(N)}.  
\end{align}
By inserting~\eqref{eq:expvalue-alpha} into~\eqref{eq:transmit-power}, and applying the power constraint in~\eqref{eq:power-constraint}, we obtain~\eqref{eq:GCB:power-constraint}.

\subsection{Proof of~\eqref{eq:SINR:parametrized}}
Let $\bg_{mk} = \hat{\bg}_{mk} + \tilde{\bg}_{mk} \in \C^N$ and $\hat{\bg}_{mk}$ be independent of $\tilde{\bg}_{mk}$. Moreover, $\bg_{mk}\!\sim\!\CN(\bzero,\beta_{mk}\bI_N)$, $\hat{\bg}_{mk}\!\sim\!\CN(\bzero,\gamma_{mk}\bI_N)$ and $\tilde{\bg}_{mk}\!\sim\!\CN(\bzero,(\beta_{mk}-\gamma_{mk})\bI_N)$. It holds that, 
\begin{align}
&\EX{\frac{ \bg_{mk}\trans \hat{\bg}_{mk}^\ast}{\norm{\hat{\bg}_{mk}}^{\alpha+1}}}\!\! = \! \EX{\!\EX{\frac{\hat{\bg}_{mk}\trans \hat{\bg}_{mk}^\ast}{\norm{\hat{\bg}_{mk}}^{\alpha+1}}+\frac{\tilde{\bg}_{mk}\trans \hat{\bg}_{mk}^\ast}{\norm{\hat{\bg}_{mk}}^{\alpha+1}}\bigg\rvert \hat{\bg}_{mk}}\!} \nonumber \\
&\quad= \EX{{\left(\norm{\hat{\bg}_{mk}}^2\right)^{\mbox{$-\frac{\alpha-1}{2}$}}}} = \frac{\Gamma(N-\frac{\alpha-1}{2})}{\Gamma(N)} \ \gamma_{mk}^{\mbox{$-\frac{\alpha-1}{2}$}},
\end{align}
which is obtained by using the same methodology as in~\eqref{eq:expvalue-alpha}. Moreover,
\begin{align}
&\EX{\frac{ |\bg_{mk}\trans \hat{\bg}_{mk}^\ast|^2}{\norm{\hat{\bg}_{mk}}^{2\alpha+2}}} = \EX{\frac{ |(\hat{\bg}_{mk} + \tilde{\bg}_{mk})\trans \hat{\bg}_{mk}^\ast|^2}{\norm{\hat{\bg}_{mk}}^{2\alpha+2}}} \nonumber \\
&\quad=\! \EX{{\left(\norm{\hat{\bg}_{mk}}^2\right)^{1-\alpha}}}+\EX{|\tilde{\bg}_{mk}\trans \hat{\bg}_{mk}^\ast|^2/\norm{\hat{\bg}_{mk}}^{2\alpha+2}} \nonumber \\
&\quad=\! \frac{\Gamma(N\!-\!\alpha\!+\!1)}{\Gamma(N)} \gamma_{mk}^{1-\alpha}\! +\! \EX{\frac{\hat{\bg}\trans_{mk}\EX{\tilde{\bg}^\ast_{mk}\tilde{\bg}\trans_{mk}}\hat{\bg}^\ast_{mk}}{\norm{\hat{\bg}_{mk}}^{2\alpha+2}}} \nonumber \\
&\quad=\! \frac{\Gamma(N\!-\!\alpha\!+\!1)}{\Gamma(N)} \gamma_{mk}^{1-\alpha}\!\!+\!(\beta_{mk}\!-\!\gamma_{mk})\EX{{\norm{\hat{\bg}_{mk}}^{-2\alpha}}}, \nonumber \\
%&\quad=\! \frac{\Gamma(N\!\!-\!\alpha\!+\!1)}{\Gamma(N)} \gamma_{mk}^{1-\alpha}\!+\!\frac{\Gamma(N\!\!-\!\alpha)}{\Gamma(N)}\gamma_{mk}^{1-\alpha}\!\left(\!\frac{\beta_{mk}}{\gamma_{mk}}\!-\!1\!\right) \nonumber \\
&\quad=\! \frac{\Gamma(N\!\!-\!\alpha)}{\Gamma(N)}\gamma_{mk}^{1-\alpha}\!\left(\!\frac{\beta_{mk}}{\gamma_{mk}}\!-\!1\!+\!N\!-\!\alpha\right),
\end{align}
where in the last step we used the identity $\Gamma(N-\alpha+1) = (N-\alpha) \Gamma(N-\alpha)$ and~\eqref{eq:expvalue-alpha}.
Consider two different UEs identified by the indices $k$ and $j$, $j \neq k$. It holds that,
\begin{align}
&\EX{\!\frac{ |\bg_{mk}\trans \hat{\bg}_{mj}^\ast|^2}{\norm{\hat{\bg}_{mj}}^{2\alpha+2}}\!}\! =\! 
\begin{cases}\!\!
\dfrac{\beta_{mk}^{2\alpha}}{\beta_{mj}^{2\alpha}} \ \EX{\dfrac{ |\bg_{mk}\trans \hat{\bg}_{mk}^\ast|^2}{\norm{\hat{\bg}_{mk}}^{2\alpha+2}}}, &\; \text{if } \bvphi_k \!=\! \bvphi_j,  \vspace*{1mm} \\
{\beta_{mk}} \ \EX{{\norm{\hat{\bg}_{mk}}^{-2\alpha}}}, &\; \text{otherwise,}
\end{cases} \nonumber \\
&\quad=\! 
\begin{cases}\!\!
\dfrac{\Gamma(N\!\!-\!\alpha)}{\Gamma(N)} \left[\dfrac{\gamma_{mk}}{\gamma_{mj}^{\alpha}} (N\!-\!1\!-\!\alpha)\!+\! \dfrac{\beta_{mk}}{\gamma^{\alpha}_{mj}}\right], &\, \text{if } \bvphi_k \!=\! \bvphi_j, \vspace*{1mm} \\
\dfrac{\beta_{mk}}{\gamma^{\alpha}_{mj}}\dfrac{\Gamma(N\!\!-\!\alpha)}{\Gamma(N)}, &\, \text{otherwise.}
\end{cases} \label{eq:expectation} \\
&\EX{\frac{ \bg_{mk}\trans \hat{\bg}_{mj}^\ast}{\norm{\hat{\bg}_{mj}}^{\alpha+1}}} \!= \!
\begin{cases} \!
\dfrac{\gamma_{mk}^{1/2}}{\gamma^{\alpha/2}_{mj}} \dfrac{\Gamma(N\!\!-\!\frac{\alpha-1}{2}\!)}{\Gamma(N)}, &\text{if } \bvphi_k\!=\! \bvphi_j,  \\
0, &\text{otherwise.}
\end{cases}
\end{align}
In these equalities, if $\bvphi_k\!\neq\!\bvphi_j$ we exploit that $\bg_{mk}$ is independent of $\hat{\bg}_{mj}$, else if $\bvphi_k\!=\!\bvphi_j$ we exploit the identities~\eqref{eq:correlated-estimates}.
By using the above results, and setting $\rho_{mk} = \Pd \eta_{mk}$, we can compute in closed form
\begin{align} 
\mathsf{DS}_k 
%&= 
%\sum_{m \in \mathcal{M}_k}\sqrt{\rho_{mk}} \ \EX{\frac{ \bg\trans_{mk} %\hat{\bg}_{mk}^\ast}{\norm{\hat{\bg}_{mk}}^{\alpha+1}}} \nonumber \\
&= \frac{\Gamma(N-\frac{\alpha-1}{2})}{\Gamma(N)}\sum_{m \in \mathcal{M}_k} \sqrt{\rho_{mk}} \gamma_{mk}^{\mbox{$-\frac{\alpha-1}{2}$}}, \label{eq:WCB:DS}
\end{align}
\begin{align} 
& \E\{\left| \mathsf{BU}_k\right|^{_2}\} \nonumber \\ 
&\quad=\! \sum_{m \in \mathcal{M}_k} \!\rho_{mk} \Bigg( \EX{\frac{|\bg\trans_{mk} \hat{\bg}_{mk}^\ast|^2}{\norm{\hat{\bg}_{mk}}^{2\alpha+2}}} - \Bigg| \EX{\frac{ \bg\trans_{mk} \hat{\bg}_{mk}^\ast}{\norm{\hat{\bg}_{mk}}^{\alpha+1}}} \Bigg|^2 \Bigg) \nonumber \\
&\quad=\! \dfrac{\Gamma(N\!-\!\alpha)}{\Gamma(N)}\!\!\! \sum_{m \in \mathcal{M}_k}\!\! \rho_{mk} \dfrac{\beta_{mk}}{\gamma^{\alpha}_{mk}}\!- \dfrac{\Gamma^2(N\!-\!\frac{\alpha-1}{2})}{\Gamma^2(N)}\!\!\!\sum_{m \in \mathcal{M}_k}\! \rho_{mk}\gamma^{1-\alpha}_{mk} \nonumber \\
&\qquad+\dfrac{\Gamma(N\!-\!\alpha)}{\Gamma(N)}(N\!-\!\alpha\!-\!1)\! \sum_{m \in \mathcal{M}_k}\! \rho_{mk}\gamma^{1-\alpha}_{mk} \label{eq:WCB:BU},
\end{align} \vspace*{-4mm}
\begin{align} 
&\EX{|\mathsf{UI}_{kj}|^2} \nonumber \\
&\quad = \sum\limits_{m \in \mathcal{M}_j} \rho_{mj} \E\bigg\{\frac{|\bg\trans_{mk} \hat{\bg}_{mj}^\ast|^2}{\norm{\hat{\bg}_{mj}}^{2\alpha+2}}\bigg\} \nonumber \\ 
&\qquad \!+\! \sum\limits_{m \in \mathcal{M}_j} \sum\limits_{\substack{n \in \mathcal{M}_j \\ n \neq m}} \sqrt{\rho_{mj} \rho_{nj}}~\E\bigg\{\frac{\bg\trans_{mk} \hat{\bg}_{mj}^\ast}{\norm{\hat{\bg}_{mj}}^{\alpha+1}}\frac{(\bg\trans_{nk} \hat{\bg}_{nj}^\ast)^\ast}{\norm{\hat{\bg}_{nj}}^{\alpha+1}}\bigg\} \nonumber \\
&\quad= \dfrac{\Gamma(N\!-\!\alpha)}{\Gamma(N)}\!\! \sum\limits_{m \in \mathcal{M}_j} \rho_{mj} \bigg[ \frac{\beta_{mk}}{\gamma^{\alpha}_{mj}} \!+\!|\bvphi\herm_k \bvphi_j|^2 \frac{\gamma_{mk}}{\gamma^{\alpha}_{mj}} (N\!-\!1\!-\!\alpha)\bigg] \nonumber \\
&\qquad \!+\! |\bvphi\herm_k \bvphi_j|^2 \dfrac{\Gamma^2(N\!-\!\frac{\alpha-1}{2})}{\Gamma^2(N)} \!\sum\limits_{m \in \mathcal{M}_j}\sum\limits_{\substack{n \in \mathcal{M}_j \\ n \neq m}}\!\!\!\sqrt{\rho_{mj} \rho_{nj}}\frac{\gamma^{1/2}_{mk}\gamma^{1/2}_{nk}}{\gamma^{\alpha/2}_{mj}\gamma^{\alpha/2}_{nj}} \nonumber \\ 
&\quad= \dfrac{\Gamma(N\!-\!\alpha)}{\Gamma(N)}\!\! \sum\limits_{m \in \mathcal{M}_j} \rho_{mj} \frac{\beta_{mk}}{\gamma^{\alpha}_{mj}}\!+\! |\bvphi\herm_k \bvphi_j|^2 \!\! \sum\limits_{m \in \mathcal{M}_j} \rho_{mj} \frac{\gamma_{mk}}{\gamma^{\alpha}_{mj}} \nonumber \\
&\qquad \times \left[ \dfrac{\Gamma(N\!-\!\alpha)}{\Gamma(N)} (N\!-\!1\!-\!\alpha)\!-\! \dfrac{\Gamma^2(N\!-\!\frac{\alpha-1}{2})}{\Gamma^2(N)} \right] \nonumber \\
&\qquad +|\bvphi\herm_k \bvphi_j|^2 \dfrac{\Gamma^2(N\!-\!\frac{\alpha-1}{2})}{\Gamma^2(N)} \Bigg(\sum\limits_{m \in \mathcal{M}_j}\!\! \sqrt{\rho_{mj}} \frac{\gamma_{mk}^{1/2}}{\gamma^{\alpha/2}_{mj}} \Bigg)^2. \label{eq:WCB:UI}  
\end{align}
%where in the last equality we added and subtracted the term $$|\bvphi\herm_k \bvphi_j|^2 \dfrac{\Gamma^2(N\!-\!\frac{\alpha-1}{2})}{\Gamma^2(N)} \sum\limits_{m \in \mathcal{M}_j} \rho_{mj} \frac{\gamma_{mk}}{\gamma^{\alpha}_{mj}}.$$
%the independence of channel responses and channel estimates of different APs ($n \neq m$), and the fact that the term $$\dfrac{\beta_{mk}}{\gamma^{\alpha}_{mj}}\dfrac{\Gamma(N\!\!-\!\alpha)}{\Gamma(N)},$$ appears regardless of pilot contamination, as shown in~\eqref{eq:expectation}. 
%Hence, this term does not depend on $|\bvphi\herm_k \bvphi_j|^2$. 
By inserting the results in~\eqref{eq:WCB:DS}--\eqref{eq:WCB:UI} into the SINR expression of~\eqref{eq:SE} we obtain~\eqref{eq:SINR:parametrized}.
\vspace*{-2mm}
\bibliographystyle{IEEEtran}
\bibliography{IEEEabrv,refs-abbr}

\end{document}